\documentclass[twocolumn, times, tighten]{aastex63}
\usepackage{textcomp}
\usepackage{amsmath}
\usepackage{color}
\usepackage{xcolor}
\usepackage{amssymb}
\usepackage{hyperref}

\shortauthors{Li et al.}
\shorttitle{AGN Accretion Disk Sizes}

\begin{document}
\nocite{*}

\title{Faint AGNs Favor Unexpectedly Long Inter-band Time Lags}

\author{Ting Li}
\affiliation{Department of Astronomy, Xiamen University, Xiamen, Fujian 
361005, China; msun88@xmu.edu.cn}

\author[0000-0002-0771-2153]{Mouyuan Sun}
\affiliation{Department of Astronomy, Xiamen University, Xiamen, Fujian 
361005, China; msun88@xmu.edu.cn}

\author{Xiaoyu Xu}
\affiliation{Department of Astronomy, Xiamen University, Xiamen, Fujian 
361005, China; msun88@xmu.edu.cn}

\author[0000-0002-0167-2453]{W. N. Brandt}
\affiliation{Department of Astronomy \& Astrophysics, 525 Davey Lab, The Pennsylvania State
University, University Park, PA 16802, USA}
\affiliation{Institute for Gravitation and the Cosmos, The Pennsylvania State
University, University Park, PA 16802, USA}
\affiliation{Department of Physics, 104 Davey Lab, The Pennsylvania State University, University 
Park, PA 16802, USA}

\author[0000-0002-1410-0470]{Jonathan R. Trump}
\affiliation{Department of Physics, University of Connecticut, Storrs, CT 06269, 
USA}

\author[0000-0003-0644-9282]{Zhefu Yu}
\affiliation{Department of Astronomy, The Ohio State University, Columbus, OH 43210, USA}

\author[0000-0002-4419-6434]{Junxian Wang}
\affiliation{CAS Key Laboratory for Research in Galaxies and Cosmology, 
Department of Astronomy, University of Science and Technology of China, 
Hefei 230026, China}
\affiliation{School of Astronomy and Space Science, University of Science 
and Technology of China, Hefei 230026, China}

\author[0000-0002-1935-8104]{Yongquan Xue}
\affiliation{CAS Key Laboratory for Research in Galaxies and Cosmology, 
Department of Astronomy, University of Science and Technology of China, 
Hefei 230026, China}
\affiliation{School of Astronomy and Space Science, University of Science 
and Technology of China, Hefei 230026, China}

\author[0000-0002-4223-2198]{Zhenyi Cai}
\affiliation{CAS Key Laboratory for Research in Galaxies and Cosmology, 
Department of Astronomy, University of Science and Technology of China, 
Hefei 230026, China}
\affiliation{School of Astronomy and Space Science, University of Science 
and Technology of China, Hefei 230026, China}

\author[0000-0003-3137-1851]{Wei-Min Gu}
\affiliation{Department of Astronomy, Xiamen University, Xiamen, Fujian 
361005, China; msun88@xmu.edu.cn}

\author[00000-0002-0957-7151]{Y. Homayouni}
\affiliation{Department of Physics, University of Connecticut, Storrs, CT 06269, 
USA}

\author[0000-0001-8678-6291]{Tong Liu}
\affiliation{Department of Astronomy, Xiamen University, Xiamen, Fujian 
361005, China; msun88@xmu.edu.cn}

\author[0000-0003-4874-0369]{Jun-Feng Wang}
\affiliation{Department of Astronomy, Xiamen University, Xiamen, Fujian 
361005, China; msun88@xmu.edu.cn}

\author[0000-0002-2419-6875]{Zhixiang Zhang}
\affiliation{Department of Astronomy, Xiamen University, Xiamen, Fujian 
361005, China; msun88@xmu.edu.cn}

\author{Hai-Kun Li}
\affiliation{Department of Astronomy, Xiamen University, Xiamen, Fujian 
361005, China; msun88@xmu.edu.cn}

\correspondingauthor{Mouyuan Sun}
\email{msun88@xmu.edu.cn}

\begin{abstract}
Inconsistent conclusions are obtained from recent active galactic nuclei (AGNs) 
accretion disk inter-band time-lag measurements. While some works show that 
the measured time lags are significantly larger (by a factor of $\sim 3$) than 
the theoretical predictions of the Shakura \& Sunyaev disk (SSD) 
model, others find that the time-lag measurements are consistent with (or only 
slightly larger than) that of the SSD model. These conflicting observational 
results might be symptoms of our 
poor understanding of AGN accretion physics. Here we show that sources with 
larger-than-expected time lags tend to be less-luminous AGNs. Such a dependence 
is unexpected if the inter-band time lags are attributed to the light-travel-time 
delay of the illuminating variable X-ray photons to the static SSD. If, instead, the 
measured inter-band lags are related not only to the static SSD but also to the 
outer broad emission-line regions (BLRs; e.g., the blended broad emission lines 
and/or diffuse continua), our result indicates that the contribution of the non-disk 
BLR to the observed UV/optical continuum decreases with increasing luminosity ($L$), 
i.e., an anti-correlation resembling the well-known Baldwin effect. Alternatively, 
we argue that the observed dependence might be a result of coherent disk thermal 
fluctuations as the relevant thermal timescale, $\tau_{\mathrm{TH}}\propto L^{0.5}$. 
With future accurate measurements of inter-band time lags, the above two 
scenarios can be distinguished by inspecting the dependence of inter-band time lags 
upon either the BLR components in the variable spectra or the timescales. 
\end{abstract}

\keywords{accretion, accretion disks---galaxies: active---quasars: general---quasars: 
supermassive black holes}

\section{Introduction}
\label{sect:intro}
Active Galactic Nucleus (AGN) continua at various UV/optical bands vary coherently, 
and the long-wavelength emission usually lags the short-wavelength emission with 
time delays of days. Such short time delays are unexpected in the static Shakura \& 
Sunyaev disk \citep[SSD;][]{SSD} model as the relevant radial propagation timescale 
(i.e., the viscous timescale) is hundreds to thousands of years. Instead, the 
inter-band cross correlations and time lags ($\tau$) are often understood in the 
framework of X-ray reprocessing \citep[e.g.,][]{Krolik1991}. In this scenario, 
the central compact X-ray corona can illuminate the disk surface and the absorbed 
X-ray emission is thermalized and reprocessed as UV/optical emission. The 
inter-band time lags account for the differences in the light-travel timescales 
from the corona to the emission regions of various UV/optical wavelengths. 

Recent high-cadence multi-band observations of several Seyfert 1 AGNs suggest that 
the measured inter-band time lags are longer than the predictions of the X-ray 
reprocessing of a static SSD by a factor of $2$ -- $3$ \citep[e.g.,][]{Fausnaugh2016, 
Cackett2018,McHardy2018, Edelson2019}. This result is further supported by 
Pan-STARRS observations \citep{Jiang2017}, which might, however, suffer 
from significant selection bias \citep[see Appendix A of][]{Homayouni2019}. 
Possible explanations involve alternative 
reprocessors, e.g., SSDs with powerful winds which have flatter disk temperature 
profiles \citep{Li2019,Sun2019}, SSDs with non-blackbody disk emission \citep{Hall2018}, 
inhomogeneous SSDs with global temperature fluctuations \citep{Cai2020, Sun2020a}, 
or non-disk UV/optical continuum emission from the more extended broad-line clouds 
\citep{Cackett2018, Lawther2018, Sun2018a, Chelouche2019,Korista2019}. 

In contrast, very recent light-curve studies of distant survey quasars (which in 
general are more luminous than the targeted studies of nearby Seyfert 1 AGNs) from 
the Dark Energy Survey \citep[DES; e.g.,][]{Flaugher2015}-Australian DES 
\citep[OzDES; e.g.,][]{Lidman2020} reverberation mapping project \citep{Yu2020}, 
the Sloan Digital Sky Survey 
Reverberation Mapping (SDSS-RM) project \citep{Homayouni2019} and the quasar 
PG 2308+098 \citep{Kokubo2018} suggest that the measured inter-band 
time lags are actually consistent with (or only slightly larger than) the 
predictions of the X-ray reprocessing of a static SSD. 

Microlensing observations of quasars find oversized disks \citep{Morgan2018}, which 
are actually inconsistent with the time-lag observations of quasars with similar 
luminosities but agree with the time lags of fainter nearby Seyfert 1 AGNs. 

These apparently conflicting observational 
results might be symptoms of our poor understanding of AGN accretion physics. 
Meanwhile, \cite{Tie2018} point out that the accretion-disk size induced microlensing 
time lags add significant systematic uncertainties to gravitational-lensing time delay 
cosmology. Hence, determining the real accretion-disk sizes are vital for our 
understanding of AGN central-engine physics and measuring cosmological parameters via 
gravitational lensing of distant quasars. 

Here, we collect a large set of inter-band time-lag measurements from different 
previously published studies to demonstrate that the ratios of the measured time 
lags to the expectations of the SSD model anti-correlate with AGN luminosity ($L$). 
That is, faint AGNs tend to have larger-than-expected inter-band time lags. This 
result is unexpected in the X-ray reprocessing of a static SSD (hereafter the 
lamp-post SSD model) and might provide critical clues to the AGN disk-size problem.

\section{The measured and predicted AGN accretion-disk sizes}
\label{sect:problem}
We use the inter-band time lags reported in previous studies to estimate the 
$\lambda = 2500\ \mathrm{\AA}$ time lags (hereafter $\tau_{\mathrm{obs}}$); 
the estimation procedures are presented in Appendix~\ref{app:1}. 
Our sample consists of the available local AGNs listed in Table~\ref{table:t1} 
and more distant quasars from DES standard star fields \citep[][hereafter Y20]{Yu2020}, 
SDSS-RM \citep[][hereafter H19]{Homayouni2019} and PG 2308+098 \citep{Kokubo2018}. 
We exclude the data from 
\cite{Jiang2017} and \cite{Mudd2018} as their relevant cadences are sparser than for 
H19 or Y20 and thus the time lags are likely biased to larger values. 

As a second step, we use the estimated black hole mass ($M_{\mathrm{BH}}$) and AGN 
bolometric luminosity ($L$) to calculate the expected 
$\lambda=2500\ \mathrm{\AA}$ static SSD lag \citep[hereafter $\tau_{\mathrm{SSD}}$; 
see Eq.~7 of][]{Fausnaugh2018}, 
\begin{equation}
  \label{eq:SSD}
  \tau_{\mathrm{SSD}} = 0.128 X (\frac{M_{\mathrm{BH}}}{10^8\ M_{\odot}})^{2/3} 
  (\frac{f_{\mathrm{Edd}}}{0.1})^{1/3}\ \mathrm{days} \\,
\end{equation}
where $f_{\mathrm{Edd}}$ is the ratio of $L$ to the Eddington luminosity. Note that 
this formula is valid only if the radiative efficiency is $\eta=0.1$; $L$ is estimated 
from the continuum luminosities at rest-frame $1350\ \mathrm{\AA}$, $3000\ \mathrm{\AA}$, 
or $5100\ \mathrm{\AA}$ (depending on redshift) with the bolometric corrections recommended 
by \cite{Richards2006}. The factor $X$ is chosen to be $5.04$; that is, $\tau_{\mathrm{SSD}}$ 
corresponds to the average time lag of the variable flux \citep{Tie2018} and is $1.5$ 
times larger than the flux-weighted light-travel time lag adopted in some previous studies 
\citep[e.g.,][]{Fausnaugh2016}. In the third step, 
the ratio ($\tau_{\mathrm{diff}}$) of the measured $\tau_{\mathrm{obs}}$ to $\tau_{\mathrm{SSD}}$ 
is obtained. 

\begin{figure}
  \epsscale{1.2}
  \plotone{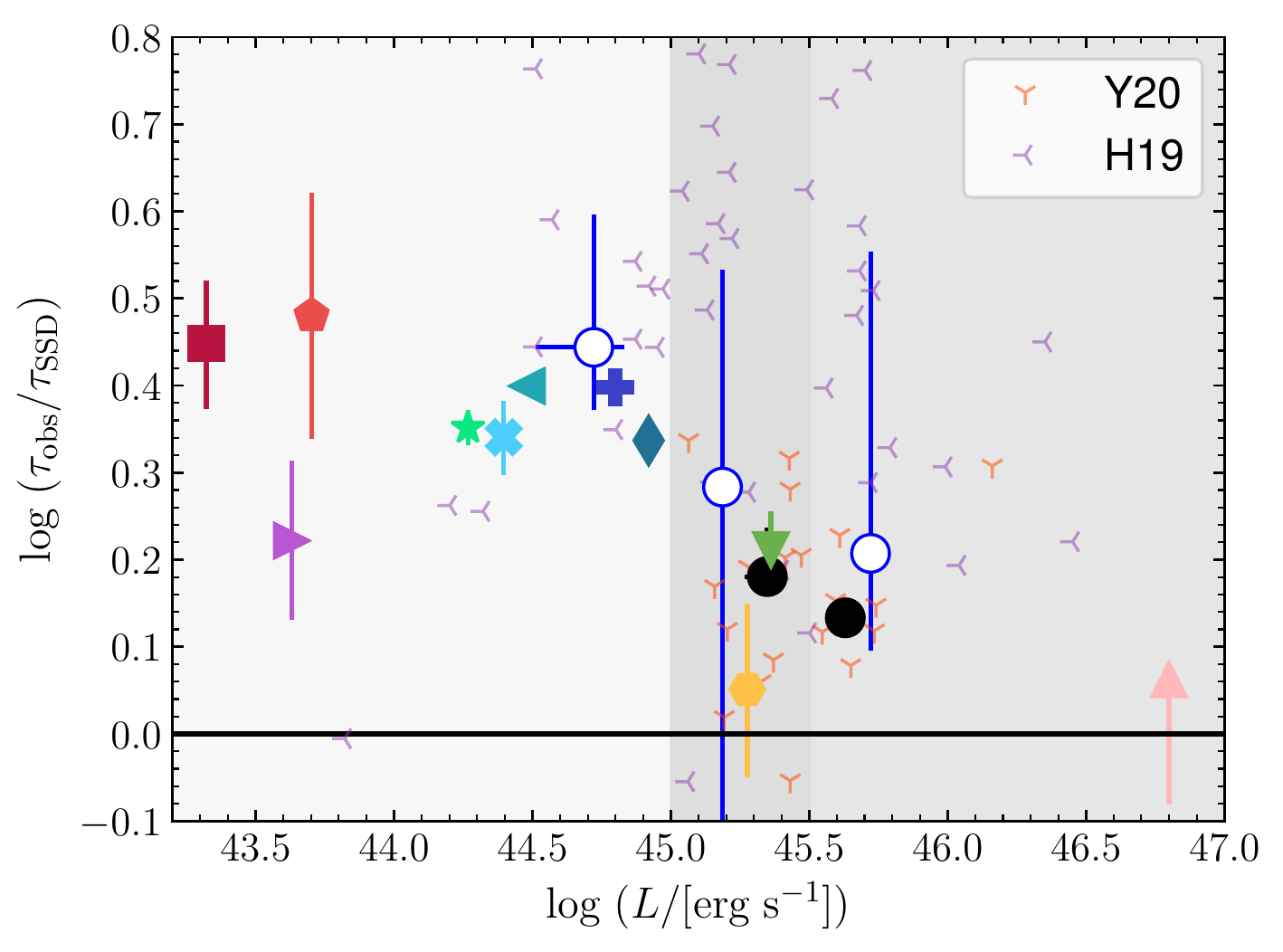}
  \caption{The ratio of the observed $\tau_{\mathrm{obs}}$ to $\tau_{\mathrm{SSD}}$ anti-correlates 
  with $L$. The large colored filled symbols with 
  error bars represent (from the lowest to highest luminosities) NGC 4593, NGC 2617, 
  NGC 4151, Mrk 142, MCG+08-11-011, NGC 5548, NGC 7469, Fairall 9, Ark 120, Mrk 509 
  and PG 2308+098, respectively; these sources usually have high-cadence and continuous 
  multi-wavelength observations. 
  The small yellow and purple symbols indicate the medians of the quasar time lags from 
  the DES (i.e., Y20) and SDSS-RM (i.e., H19) surveys (while all H19 sources are taken 
  into account in the medians and all subsequent analysis, some H19 sources with negative, 
  extremely small or large ratios are not presented in this figure for the sake of clarity). 
  The H19 sources are divided 
  into three luminosity bins (i.e., $\log (L/[\mathrm{erg\ s^{-1}}])<45$, 
  $45<\log (L/[\mathrm{erg\ s^{-1}}])<45.5$ and $\log (L/[\mathrm{erg\ s^{-1}}]) 
  >45.5$, respectively; see also the shaded gray regions) and the three bins have 
  similar source numbers; the DES AGNs are divided into two luminosity bins 
  ($\log (L/[\mathrm{erg\ s^{-1}}])<45.5$ and $\log (L/[\mathrm{erg\ s^{-1}}])>45.5$, 
  respectively). The blue open and black filled circles with error bars represent the 
  median ratios for these luminosity bins and their $1\sigma$ uncertainties (estimated 
  via bootstrapping; the error bars of the black filled circles are too small to show). 
  \label{fig:fobs}}
\end{figure}

We find that $\tau_{\mathrm{diff}}$ anti-correlates with $L$ (see Figure~\ref{fig:fobs}), 
with a Spearman correlation coefficient $\rho_{\mathrm{spearman, obs}} 
= -0.78$ (the corresponding $p$-value is $3\times 10^{-4}$).\footnote{We use the local-AGN 
measurements and the median values of the sources from H19 and Y20 to calculate the 
Spearman coefficient unless otherwise specified. This is because, comparing with local 
AGN measurements, the results of H19 and Y20 have larger uncertainties; thus, we decide 
to use their median values. Nevertheless, we also calculate the corresponding Spearman 
coefficient by treating the local AGNs and the quasars from H19 and Y20 equally and 
find that the coefficient and the $p$-value are $-0.29$ and $3\times 10^{-3}$, respectively; 
that is, the anti-correlation still exists.} We also confirm that $\tau_{\mathrm{diff}}$ 
anti-correlates with $M_{\mathrm{BH}}$ (the Spearman coefficient and the $p$-value are 
$-0.69$ and $4\times 10^{-3}$, respectively); however, the partial correlation 
between $\tau_{\mathrm{diff}}$ and $M_{\mathrm{BH}}$ is insignificant once $L$ is 
controlled (the Spearman partial-correlation coefficient, which is calculating via the 
package ``ppcor'' in R, is $-0.05$). Our results also 
suggest an apparent disagreement between the microlensing disk-size measurements and 
the inter-band time-lag observations: the former focus on luminous AGNs but find 
over-sized disks and the latter suggest that such AGNs have consistent-with-expectation 
time lags. 

To test whether the lamp-post SSD model can 
explain our results or not, we perform the following experiment. First, we use the 
measured $M_{\mathrm{BH}}$ and $L_{\mathrm{bol}}$ to calculate the static SSD 
effective temperature ($T_{\mathrm{eff}}$) profile (we again adopt $\eta=0.1$). 
Second, we add the additional surface heating due to the X-ray illumination (which 
is modeled by a damped random-walk process) to obtain the fluctuations of 
$T_{\mathrm{eff}}$ and the corresponding mock UV/optical light curves (by integrating 
the SSD blackbody emission over the whole disk). 
Note that, the adopted UV/optical bands are the same as real observations. Third, the 
mock UV/optical inter-band time lags are estimated by utilizing \textit{Javelin} to 
fit the mock light curves. Fourth, we calculate the corresponding mock time lags 
at $2500\ \mathrm{\AA}$ (hereafter $\tau_{\mathrm{SSD(sim)}}$) following 
the same recipes aforementioned. We repeat this experiment $200$ times to account 
for statistical fluctuations. 

\begin{figure*}
  \epsscale{1.2}
  \plotone{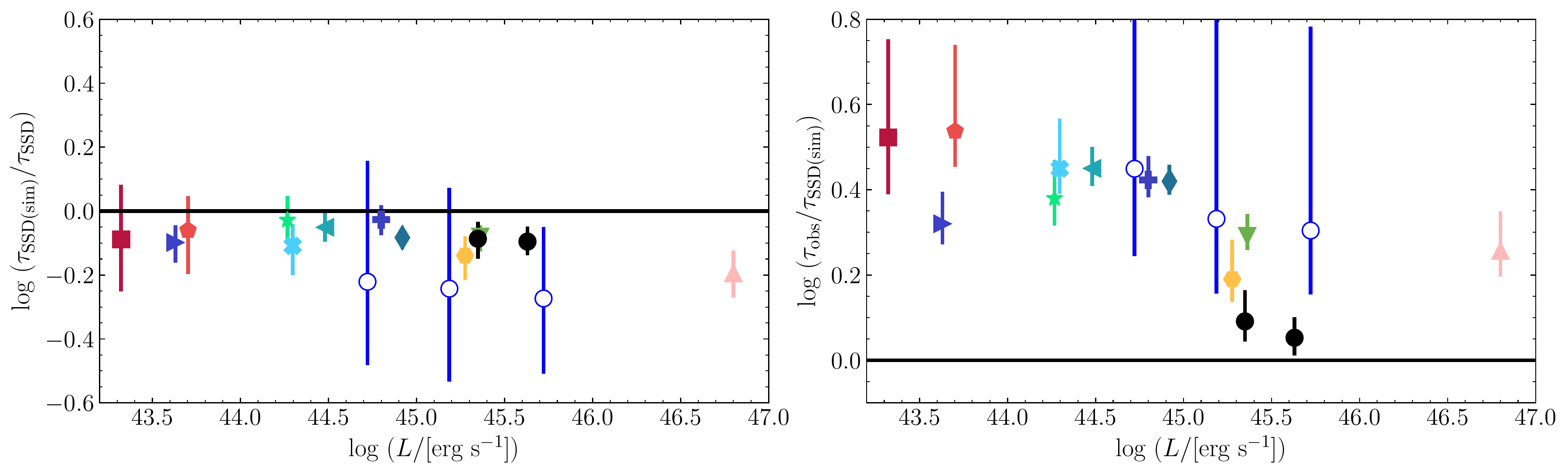}
  \caption{Left-panel: the ratio of $\tau_{\mathrm{SSD(sim)}}$ (i.e., obtained by applying 
  \textit{Javelin} to analysis the lamp-post SSD simulations that mimic real observations) 
  to $\tau_{\mathrm{SSD}}$ (i.e., Eq.~\ref{eq:SSD}) versus $L$. Right-panel: the 
  ratio of $\tau_{\mathrm{obs}}$ to $\tau_{\mathrm{SSD(sim)}}$ 
  versus $L$. The symbols share the same meanings as in Figure~\ref{fig:fobs}. 
  The lamp-post SSD model cannot explain our results. \label{fig:SSD}}
\end{figure*}

The median ratios (and their $1\sigma$ uncertainties) of $\tau_{\mathrm{SSD(sim)}}$ to 
$\tau_{\mathrm{SSD}}$ (i.e., Eq.~\ref{eq:SSD}) are shown in the left panel of 
Figure~\ref{fig:SSD}, which suggests 
that $\tau_{\mathrm{SSD(sim)}}$ is on average less than $\tau_{\mathrm{SSD}}$; similar 
results have been reported by \cite{Chan2020} who therefore propose that the 
accretion-disk sizes estimated via \textit{Javelin} are underestimates by $\sim 30\%$ 
(or $0.15$ dex). However, such a bias cannot account for our results. Indeed, the anti-correlation 
in the left panel of Figure~\ref{fig:SSD} is statistically insignificant (the Spearman 
coefficient and the $p$-value are $-0.38$ and $0.14$, respectively) and much weaker than 
the observed one. Moreover, a clear anti-correlation between the ratios of 
$\tau_{\mathrm{obs}}$ to $\tau_{\mathrm{SSD(sim)}}$ and $L$ holds (see the 
right panel of Figure~\ref{fig:SSD}; the Spearman coefficient and the $p$-value are 
$-0.77$ and $5\times 10^{-4}$, respectively). 

\section{Physical Implications}
\label{sect:imp}
Our results are unexpected in the lamp-post SSD model. Some 
alternative solutions are proposed to account for the AGN disk-size problem. Below, 
we discuss the physical implications of our results for these models. 
\subsection{Implications for the X-ray reprocessing models}
\label{sect:xray}
One model to explain the larger-than-expected time lags for Seyfert AGNs involves 
disk-atmosphere radiative transfer effects \citep{Hall2018}. If so, our results 
suggest that the atmosphere effects are weaker in more luminous AGNs. 
It is unclear what mechanisms can drive this behavior. 

Alternatively, \cite{Sun2019} \citep[also see][]{Li2019} consider SSDs with 
powerful winds and find that such disks can have larger apparent sizes than the 
no-wind SSDs. Then, our results indicate that the disk-wind strength should 
decrease with increasing $L$, contradicting observations that luminous AGNs 
generally have stronger disk winds than their faint counterparts 
\citep[e.g.,][]{Laor2002, Ganguly2007}. Therefore, it seems that the lamp-post 
wind SSD is not the general solution to the disk-size problems of AGNs with 
various luminosities although winds might still play an important role in 
luminous AGNs (see Section~\ref{sect:micro}). 

The discrepancy between the observed UV/optical time lags and the lamp-post 
SSD ones might also be reconciled by increasing the corona scale height to $\gtrsim 20$ 
Schwarzschild radii \citep{Kammoun2021}, which effectively enlarges the light 
travel time. To explain our results, the corona height must then anti-correlate with 
luminosity. 

Instead, several works \citep{Cackett2018,Lawther2018,Sun2018a,Chelouche2019,Korista2019} 
propose that the diffuse nebular emission from the more distant broad-line region 
(BLR) clouds also acting as a reprocessor can produce non-disk UV/optical continuum light 
curves; then, the measured time lags are superpositions of the disk and BLR light-travel 
time delays and are thus longer than the SSD expectations. This BLR model might explain 
our results if its contribution to the UV/optical continuum emission anti-correlates with 
$L$. There is a well-known anti-correlation between broad-line strength and $L$, a.k.a., 
the Baldwin effect \citep{Baldwin1977}, and this would imply that less-luminous AGNs 
similarly have stronger diffuse nebular emission from the BLR. In other 
words, faint AGNs tend to have larger ratios of the observed to SSD model time lags. 
We stress that detailed BLR calculations should be performed to quantitatively test this 
model against our results because the slope of the Baldwin effect for various lines are 
generally not steep \citep[i.e., the line equivalent width $\propto L^{-\gamma}$ with 
$\gamma \sim 0.1$; see, e.g.,][]{Shields2007}. Such calculations are beyond the scope 
of this work; the BLR diffuse nebular emission models, e.g., \cite{Korista2019}, 
might be expanded to search for such an effect. 

We point out that these X-ray reprocessing models cannot simultaneously explain 
several other aspects of AGN UV/optical variability, 
e.g., the timescale-dependent color variations \citep[i.e., the 
lamp-post SSD can explain the variable spectra, but cannot account for the fact 
that the color variations are timescale dependent; see, e.g.,][]{Zhu2018} and the 
anti-correlations between variability amplitude and $L$ \cite[e.g.,][]{MacLeod2010, 
Sun2018b}. Below, we discuss the alternative disk-corona magnetic coupling 
scenario \citep{Sun2020a}, which has shown to be successful in reproducing the 
color variations and the dependences of the variability amplitude upon $L$ 
\citep{Sun2020b}, to understand our results in Figure~\ref{fig:fobs}.

\subsection{Implications for the MHD accretion physics}
\label{sect:char}
Theoretically speaking, disk temperatures ($T$) should vary in response to the time-dependent 
magnetohydrodynamic (MHD) turbulent heating with a response on the thermal timescale 
($\tau_{\mathrm{TH}}$). For an 
SSD, $\tau_{\mathrm{TH}}\sim \alpha^{-1}(GM_{\mathrm{BH}}R^{-3})^{-1/2}$, 
where $\alpha$, $G$, and $R$ are the dimensionless viscosity parameter, the 
gravitational constant, and the distance to the central black hole, respectively. Therefore, 
the short-wavelength emission which is produced by the inner hotter plasma has a smaller 
response thermal timescale than the long-wavelength emission. The differences in the 
response thermal timescale would add an additional inter-band time lag ($\tau_{\mathrm{add}}$), 
beyond the light-travel time lags. On short timescales ($t_{\mathrm{dur}}$), the observed 
UV/optical variations are mainly produced in the similar small-$R$ regions with 
$\tau_{\mathrm{TH}} \leq t_{\mathrm{dur}}$ as the gas in these regions can vary its temperature 
significantly, and $\tau_{\mathrm{add}}$ is small and negligible. And vice versa. For an SSD, 
at the characteristic radius $R_{\lambda}$ 
(with $k_{\mathrm{B}}T(R_{\lambda})=hc/\lambda$, where $k_{\mathrm{B}}$ and $h$ 
are the Boltzmann and Planck constants, respectively; and $T(R_{\lambda})\propto 
(M_{\mathrm{BH}}\dot{M})^{1/4}R_{\lambda}^{-3/4}$), $\tau_{\mathrm{TH}}(R_{\lambda})$ 
scales as $\alpha^{-1}L^{0.5}\lambda^2$. Therefore, $\tau_{\mathrm{add}}$ would only 
be prominent for AGNs with small $L$ (i.e., short $\tau_{\mathrm{TH}}(R_{\lambda})$) or 
large $t_{\mathrm{dur}}$. Below, we consider a specific thermal-fluctuation 
model to qualitatively reproduce the observed result. 

\begin{figure*}
    \epsscale{1.2}
    \plotone{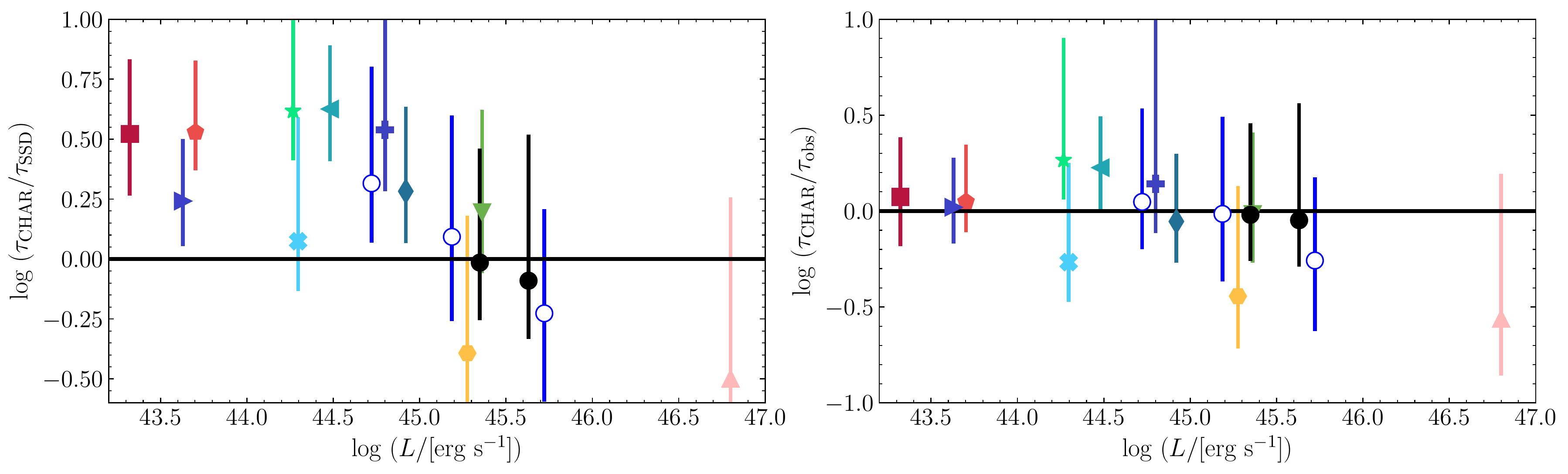}
    \caption{Left-panel: the ratio of $\tau_{\mathrm{CHAR}}$ to $\tau_{\mathrm{SSD}}$ (i.e., 
    Eq.~\ref{eq:SSD}) versus $L$. Right-panel: the ratio of $\tau_{\mathrm{CHAR}}$ to 
    $\tau_{\mathrm{obs}}$ versus $L$. The symbols share the same meanings as in Figure~\ref{fig:fobs}. 
    The CHAR model (on average) successfully accounts for the observed inter-band time lags. 
    \label{fig:char}}
\end{figure*}

In earlier work, we proposed the corona-heated accretion-disk reprocessing model (hereafter 
the CHAR model) by assuming the accretion disk and the extremely compact X-ray corona 
are magnetically coupled \citep{Sun2020a}. Magnetic fluctuations in the corona can alter 
the disk surface magnetic field, drive variations in the disk MHD 
turbulent dissipation, and eventually induce coherent multi-band variations. 
We have predicted that $\tau_{\mathrm{dif}}$ increases with decreasing $L$ for sources with 
similar monitoring durations\citep[see Figures 8 \& 16 of][]{Sun2020a}. 

The first two model parameters, $M_{\mathrm{BH}}$ and $\dot{M}$, are 
fixed according to the observational constraints. The only free parameter $\alpha$ might 
be $0.1$ -- $0.4$ according to studies of the outbursts of dwarf nova or soft X-ray 
transients \citep[e.g.,][]{King2007}. We fix\footnote{Note that the CHAR model with 
$\alpha \simeq 0.2$ can also explain the results.} $\alpha =0.1$ and obtain the mock 
multi-wavelength light curves by considering the CHAR model for all sources in Figure~\ref{fig:fobs}. 
The mock inter-band time lags are measured by again adopting \textit{Javelin}. 
The ratios (hereafter $\tau_{\mathrm{diff, mock}}$) of the CHAR time lags (hereafter 
$\tau_{\mathrm{CHAR}}$) and $\tau_{\mathrm{SSD}}$ are calculated following the same recipes 
aforementioned. 
Indeed, $\tau_{\mathrm{diff, mock}}$ also anti-correlates with $L$ (see the left panel of 
Figure~\ref{fig:char}). Moreover, the ratios of $\tau_{\mathrm{CHAR}}$ to $\tau_{\mathrm{obs}}$ 
show a lack of dependence upon $L$ (see the right panel of Figure~\ref{fig:char}), i.e., the 
CHAR model can explain our results. 

For an AGN, our CHAR model also predicts that the inter-band time lags of the slow 
variations are longer than those of the fast variations \citep[see Figure 7 of][]{Sun2020a}. 
Recently, instead of simply measuring the inter-band time lags of NGC 7469 (which is 
considered in Figure~\ref{fig:fobs}), \cite{Pahari2020} further 
split the $1825\ \mathrm{\AA}$ and $1315\ \mathrm{\AA}$ light 
curves into slow (i.e., timescale $> 5$ days) and fast (timescale $< 5$ days) 
variations and find that the corresponding time lags are $0.29\pm 0.06$ days and $0.04\pm 0.12$ 
days, respectively. We fix the two CHAR model parameters, 
$M_{\mathrm{BH}}$ and $L$, according to the NGC 7469 observations and $\alpha=0.1$ to generate 
the mock light curves. The mock light curves are also split into slow and fast components 
following the methodology of \cite{Pahari2020}. The resulting lag for the slow (fast) variation 
component is  $0.27\pm 0.2$ days ($0.04\pm 0.03$ days) which agrees with the results by 
\cite{Pahari2020}. The anti-correlation between the time lag and the timescale might also be 
responsible for the fact that NGC 4593 and NGC 4151 do 
not seem to follow the anti-correlation found in Figure~\ref{fig:fobs}. Indeed, while NGC 4593 
and NGC 4151 have the lowest luminosities, their size ratios are not the largest. We speculate 
that this is because their monitoring time durations are relatively short ($\lesssim 60$ days). 

\section{Comparing with microlensing observations}
\label{sect:micro}
Our results in Figure~\ref{fig:fobs} suggest that microlensing observations of quasars 
(which often find larger-than-expected disk sizes) are, in fact, inconsistent with the time-lag 
observations of quasars with similar luminosities. Note that the inter-band time lags of a 
gravitationally lensed AGN 0957+561 were obtained by \cite{Gil-Merino2012}. By 
interpreting the time lags as the variability-weighted ones (i.e., Eq.~\ref{eq:SSD} with $X=5.04$), 
we use their UV/optical measurements and the methodology aforementioned to obtain the corresponding 
$r$-band half-light radius (i.e., $\log (r_{\mathrm{half}}(r)/\mathrm{cm}) = 15.82$). Recently, 
\cite{Cornachione2020} obtained its microlensing $\log (r_{\mathrm{half}}(r)/\mathrm{cm}) = 
16.66^{+0.37}_{-0.62}$. The difference of the two results is $0.84$ dex albeit with 
a large uncertainty. Hence, any X-ray 
reprocessing models that actually enlarge the effective disk sizes might not explain 
this discrepancy. 

In our CHAR model, the actual disk sizes are similar to the SSD model. In addition, 
the median time lag of our CHAR model is considerably smaller than the SSD model (see 
the left panel of Figure~\ref{fig:char}). Disk winds are probably common in luminous AGNs 
\citep[e.g.,][]{Laor2002, Ganguly2007, Laor2014} and can make the actual disk sizes 
larger than the no-wind SSD model \citep[e.g.,][]{Sun2019}. If we also consider the CHAR model 
with winds, the predicted median time lag approaches the SSD model and is more consistent with 
observations. At the same time, the windy CHAR model has a larger half-light radius and can 
account for the microlensing observations \citep[e.g.,][]{Morgan2018,Li2019}. In summary, 
the possible discrepancy between the inter-band time lags and microlensing observations might 
be a unique probe of winds and corona-disk magnetic coupling. Future inter-band time-lag 
measurements of gravitationally lensed quasars can verify this idea.

\section{Summary}
We have collected the inter-band time lags for a large AGN sample. We find 
tentative evidence that the ratio of the observed to SSD time lags anti-correlates with 
$L$; this anti-correlation is unexpected in the lamp-post SSD (with or without 
winds and/or disk-atmosphere scattering) model. Our result indicates that the inter-band 
time lags are not solely determined by the spatial sizes of the emission regions but contain 
important information regarding the disk inner thermal fluctuations or the BLR structure. 
While both the BLR and CHAR models can explain our results, they have entirely 
different predictions. For the BLR model, we would expect that the time-lag ratios depend 
upon the BLR components in the variable spectra. For the CHAR model, we suggest that 
the time lags increase with the variability timescales. Future time-domain surveys such 
as the Rubin Observatory's Legacy Survey of Space and Time 
\citep[e.g.,][]{LSST} can measure inter-band time lags for a large number of AGNs with 
various properties, thereby starting a new era in testing MHD accretion-disk theory or BLR 
physics. 

\acknowledgments 
We thank the anonymous referee for his/her useful and timely comments that improved the 
manuscript. 
M.Y.S. acknowledges support from the National Natural Science Foundation of China 
(NSFC-11973002). W.N.B. acknowledges support from NSF grant AST-1516784 and NASA ADAP grant 
80NSSC18K0878. J.R.T. acknowledges support from NASA grants HST-GO-15260, HST-GO-15650, 
and 18-2ADAP18-0177, and NSF grant CAREER-1945546. 
Z.F.Y. was supported in part by the United States National Science Foundation under Grant No. 
161553. 
Y.Q.X. and J.X.W. acknowledge support from the National Natural Science Foundation of China 
(NSFC-12025303, 11890693, 11421303), the CAS Frontier Science Key Research Program 
(QYZDJ-SSW-SLH006), and the K.C. Wong Education Foundation. 
Z.Y.C. acknowledges support from the National Natural Science Foundation of China (NSFC-11873045). 
J.X.W., W.M.G., Z.Y.C. and Z.X.Z. acknowledge support from the National National Science Foundation 
of China (NSFC-12033006). 
T.L. acknowledges support from the National Natural Science Foundation of China (NSFC-11822304). 
Y.H. acknowledges support from NASA grants HST-GO-15650. 
J.F.W. acknowledges support from National Natural Science Foundation of China (NSFC-U1831205). 
We thank Prof. Jianfeng Wu for his helpful suggestions. 

\clearpage
\appendix
\section{Estimating the time lags at $2500\ \mathrm{\AA}$}
\label{app:1}
The time-lag measurements of previous studies correspond to various rest-frame 
wavelengths. Hence, for each source, we fit the function 
$\tau=\tau_0 ((\lambda/\lambda_0)^{\beta} - 1)$ with $\beta\equiv 4/3$ (as $\beta$ cannot 
be well constrained for most sources) to these time-lag measurements by minimizing the 
$\chi^2$ statistic (i.e., with only one free parameter $\tau_0$), where $\tau$ and 
$\lambda$ are the rest-frame time lag and wavelength (the subscript $0$ indicates 
the reference band), respectively. 
The best-fitting $\tau_0$ is used to infer $\tau_{\mathrm{obs}}$ (i.e., 
$\tau_{\mathrm{obs}}=\tau_0 (2500\ \mathrm{\AA}/\lambda_0)^{\beta}$). Note that the 
cross correlation and \textit{Javelin} \citep{Zu2011} are often used to measure the 
time lags of local AGNs; for more distant luminous AGNs, \textit{Javelin} is often 
preferred. We adopt the \textit{Javelin} time-lag measurements reported by previous 
studies to fit the above function; if the \textit{Javelin} time-lag measurements 
are unavailable, we use the cross-correlation centroids (e.g., NGC 4151) since 
the two methods are generally consistent with each other \citep[see, e.g., Figure 4 
of][]{Fausnaugh2016}. We note that the X-ray bands are excluded when fitting 
the lag-wavelength function. The $u$-band is also excluded except for the $z=0.433$ 
quasar, PG2308+098 (for this source, the $u$-band is adopted as the reference band 
and the corresponding rest-frame wavelength is $2478\ \mathrm{\AA}$). The adopted 
bands for each source are listed in Table~\ref{table:t1}. 
\begin{deluxetable*}{c|c|c|c}
  \tablecaption{The adopted bands to fit the inter-band lag-wavelength relation. \label{table:t1}}
  \tablenum{1}
  \tablehead{\colhead{Source name} & \colhead{Adopted bands} & \colhead{Reference band} & \colhead{Reference}} 
  
  \startdata
  NGC4593 & \begin{tabular}[c]{@{}l@{}}
    UVM2, UVW1, UVB, UVV, 1150 $\mathrm{\AA}$,\\ 1350 $\mathrm{\AA}$, 1460 $\mathrm{\AA}$, 1690 $\mathrm{\AA}$, 
    4745 $\mathrm{\AA}$,\\ 5100 $\mathrm{\AA}$, 5450 $\mathrm{\AA}$, 5600 $\mathrm{\AA}$, 6250 $\mathrm{\AA}$,\\ 
    6850 $\mathrm{\AA}$, 7450 $\mathrm{\AA}$, 8000 $\mathrm{\AA}$, 8800 $\mathrm{\AA}$, 9350 $\mathrm{\AA}$ 
  \end{tabular} & UVW2 & \cite{Cackett2018} \\
  \hline
  NGC 2617 & UVW2, UVM2, UVW1, UVB, g, UVB, r, i, z & 5100 $\mathrm{\AA}$ & \cite{Fausnaugh2018} \\
  \hline
  MCG+08-11-011 & r, i, z & g & \cite{Fausnaugh2018} \\
  \hline
  NGC 4151 & UVM2, UVW1, UVB, UVV & UVW2 & \cite{Edelson2017} \\
  \hline
  Mrk 142 & UVM2, UVW1, UVB, g, V, UVV, r, I, z & UVW2 & \cite{Cackett2020} \\
  \hline
  NGC 5548 & \begin{tabular}[c]{@{}l@{}} 
    1158 $\mathrm{\AA}$, 1479 $\mathrm{\AA}$, 1746 $\mathrm{\AA}$, UVW2, UVM2, UVW1,\\ B, UVB, g, V, r, R, I, i, z \end{tabular} 
    & 1367 $\mathrm{\AA}$ & \cite{Fausnaugh2016} \\
  \hline
  Fairall 9 & UVM2, UVW1, UVB, UVV, B, g, v, r, i, $z_{\rm s}$ & UVW2 & \cite{Santisteban2020} \\
  \hline
  Ark 120 & B, UVW1, UVB, UVV & I & \cite{Lobban2020} \\
  \hline
  NGC 7469 & V & UVW2 & \cite{Pahari2020} \\
  \hline
  Mrk 509 & UVM2, UVW1, UVB, UVV & UVW2 & \cite{Edelson2019} \\
  \hline
  PG 2308+098 & g, r, i, z & u & \cite{Kokubo2018} \\
  \hline
  H19 quasars & i & g & \cite{Homayouni2019} \\
  \hline
  Y20 quasars & r, i, z & g & \cite{Yu2020} \\
  \hline
  \enddata
  
  
  
  
  \end{deluxetable*}

\end{document}